\documentclass[journal,twoside]{IEEEtran}

\usepackage{amsmath,amsfonts,amssymb,bm}

\newcommand{\Hline}{\noalign{\hrule height 0.4mm}}

\newcommand{\ie}{\textit{i.e.}}
\newcommand{\eg}{\textit{e.g.}}

\newcommand{\etal}{\textit{et~al.}}

\newcommand{\mA}{\mathbf{A}}

\newcommand{\mS}{\mathbf{S}}

\newcommand{\mX}{\mathbf{X}}
\newcommand{\mXalt}{\mathbf{X}}

\newcommand{\mZ}{\mathbf{Z}}

\newcommand{\vs}{\mathbf{s}}

\newcommand{\vx}{\mathbf{x}}

\newcommand{\vz}{\mathbf{z}}

\usepackage{amsmath,amsfonts,amssymb,bm}
\usepackage{graphicx,url,times}
\usepackage{hhline}
\usepackage{booktabs}
\usepackage{color}
\usepackage{cite}

\allowdisplaybreaks[4]

\def\defeqn{\stackrel{\triangle}{=}}

\begin{document}
\title{Semi-supervised Neural Chord Estimation\\
Based on a Variational Autoencoder\\
with Latent Chord Labels and Features}

\author{Yiming~Wu,
Tristan~Carsault,
Eita~Nakamura,
Kazuyoshi~Yoshii,~\IEEEmembership{Member,~IEEE}
\thanks{
Manuscript 
received March 18, 2020; 
revised  XXXX XX, 2020;
accepted XXXX XX, 2020; 
Date of publication XXXX XX, 2020; 
date of current version XXXX XX, 2020.
This work was supported by JST ACCEL No. JPMJAC1602
 and JSPS KAKENHI Nos. 19H04137, 19K20340, and 16H01744.
The associate editor coordinating the review of this manuscript 
 and approving it for publication is XXX XXX. 
 \textit{(Corresponding author: Yiming Wu.)}
}
\thanks{
Y. Wu, E. Nakamura, and K. Yoshii
are with Graduate School of Informatics, Kyoto University, Kyoto 606-8501, Japan
(email: \{wu, enakamura, yoshii\}@sap.ist.i.kyoto-u.ac.jp).
E. Nakamura is also with Hakubi Center for Advanced Research, Kyoto University, Kyoto 606-8501, Japan.
}
\thanks{
T. Carsault is with Repr\'{e}sentations Musicales, IRCAM, 75004 Paris, France
(email: tristan.carsault@ircam.fr).
}
\thanks{Digital Object Identifier 10.1109/TASLP.2020.XXXXXXX}
}

\markboth{IEEE/ACM TRANSACTIONS ON AUDIO, SPEECH, AND LANGUAGE PROCESSING, VOL. XX, NO. XX, XXXX 2020}
{Yiming \MakeLowercase{\textit{et al.}}: Semi-supervised Neural Chord Estimation}

\maketitle

\begin{abstract}
This paper describes a statistically-principled semi-supervised method 
 of automatic chord estimation (ACE)
 that can make effective use of music signals 
 regardless of the availability of chord annotations.
The typical approach to ACE is to train a deep \textit{classification} model 
 (neural chord estimator) in a supervised manner
 by using only annotated music signals.
In this \textit{discriminative} approach,
 prior knowledge about chord label sequences (model output) 
 has scarcely been taken into account. 
In contrast,
 we propose a unified \textit{generative} and \textit{discriminative} approach
 in the framework of amortized variational inference.
More specifically, 
 we formulate a deep \textit{generative} model
 that represents the generative process 
 of chroma vectors (observed variables)
 from discrete labels and continuous features (latent variables),
 which are assumed to follow
 a Markov model favoring self-transitions
 and a standard Gaussian distribution, respectively.
Given chroma vectors as observed data,
 the posterior distributions of the latent labels and features
 are computed approximately 
 by using deep \textit{classification} and \textit{recognition} models, respectively.
These three models form a variational autoencoder
 and can be trained jointly in a semi-supervised manner.
The experimental results show that 
 the regularization of the classification model
 based on the Markov prior of chord labels and the generative model of chroma vectors
 improved the performance of ACE even under the supervised condition.
The semi-supervised learning using additional non-annotated data
 can further improve the performance.
\end{abstract}


\fussy

\IEEEpeerreviewmaketitle

\section{Introduction}

\IEEEPARstart{A} chord is a mid-level representation 
 of polyphonic music that lies between the 
 highly-abstracted musical intentions of humans
 and actual musical sounds.
Unlike musical notes,
 chord sequences do not tell the actual pitches,
 but abstractly represent the harmonic content evolving over time.
In lead sheets
 (a form of music notation consisting of melody, chords, and lyrics),
 chord labels are shown to musicians to roughly suggest 
 the intentions of how musical notes should be arranged and played in each bar.

Automatic chord estimation (ACE), which aims to 
 recognize the chord sequence behind a music signal,
 has been one of the fundamental research topics
 in the field of music information retrieval (MIR).
For example, 
 it forms the basis of music content visualization 
 (\eg, Songle~\cite{Goto2011}),
 and higher-level MIR tasks
 such as genre classification~\cite{anglade2009genre}
 and cover song retrieval~\cite{bello2007audio}.
The diversity and complexity
 of the acoustic characteristics of music signals
 make the ACE task very challenging.
ACE studies have thus been focusing on
 data-driven methods based on statistical models\cite{Pouwels2019},
 which are roughly categorized into
 \textit{generative} and \textit{discriminative} approaches.

\begin{figure}[t]
 \centering
 \centerline{\includegraphics[width=.95\columnwidth]{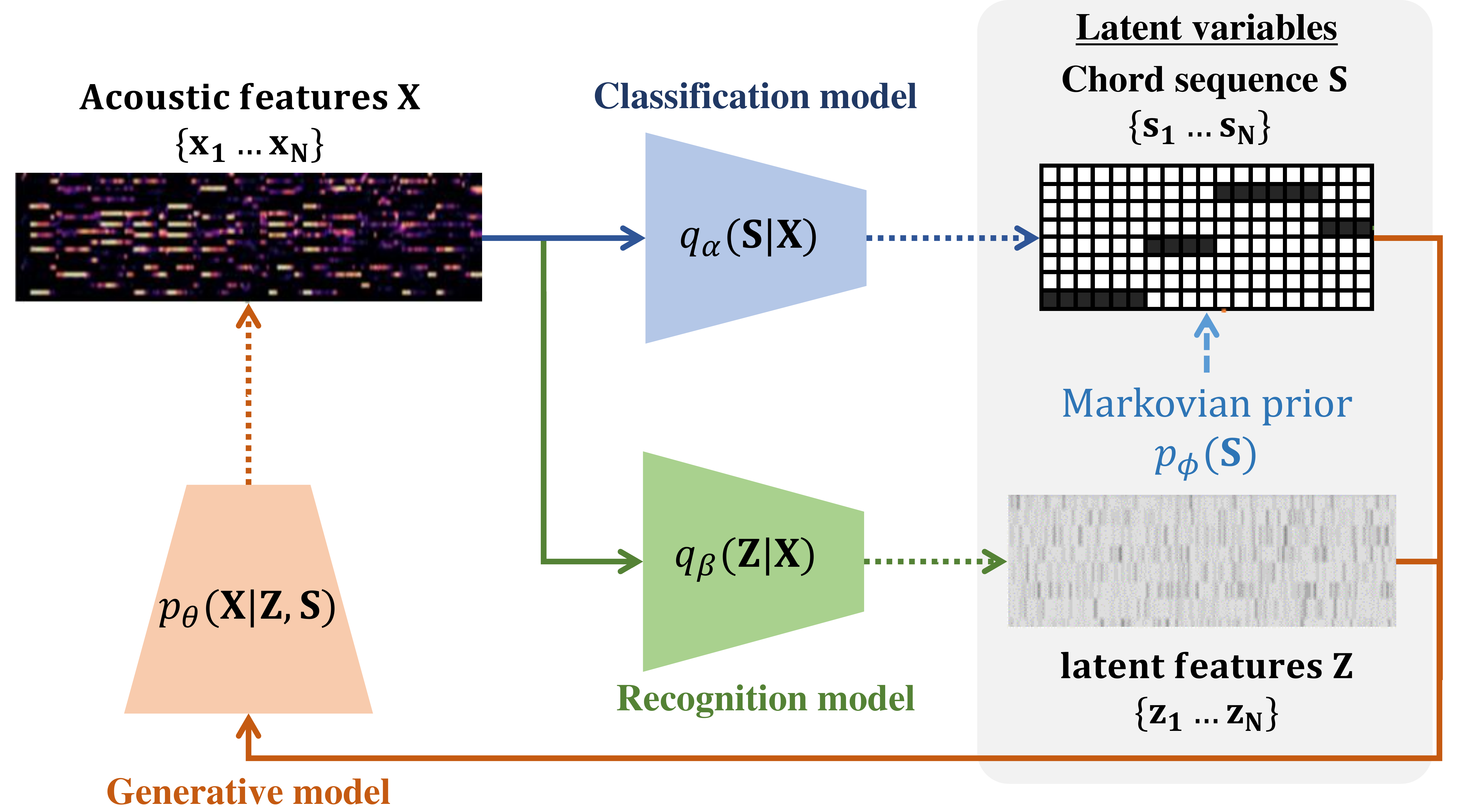}}
 \vspace{-1mm}
 \caption{The proposed variational autoencoder
 consisting of a deep generative model of chroma vectors,
 a deep classification model of chord labels, 
 and a deep recognition model of latent features.
 Dashed arrows indicate stochastic relations.
 These three models are trained jointly in a semi-supervised manner
 by using annotated and non-annotated music signals.}
 \label{fig:overview}
 \vspace{-2mm}
\end{figure}

Early studies took the \textit{generative} approach 
 based on a classical probabilistic latent variable model.
For example, a hidden Markov model (HMM) is often used
 to represent the stochastic generative process 
 of chroma vectors (observed variables) 
 from chord labels (latent variables) \cite{Lee2008}.
Given chroma vectors, 
 the chord labels are inferred
 such that their likelihood is maximized.
An advantage of this approach
 is that the prior distribution of chord labels (language model)
 can be naturally introduced.
To encourage the estimated chord labels to be consistent with musical grammar,
 the temporal continuity and dynamics of chord labels should be considered.
Another advantage is that 
 in theory generative models can be trained 
 in a semi-supervised or even unsupervised manner.
However,
 the expressive power of simple generative models
 like HMMs have been insufficient to represent
 the temporal structure of chord labels and
 the relationships between acoustic features and chord labels.

The \textit{discriminative} approach, in contrast,
 aims to directly convert acoustic features into chord labels.
Recently, deep neural networks (DNNs) have often been used 
 for estimating the posterior probabilities of chord labels. 
Because the generative process is not considered explicitly, 
 the oversimplified assumptions 
 posed by the generative approach
 can be avoided\cite{McVicar2014}, 
 and chord label inference is more straightforward.
A problem here is that
 the classifier is usually trained for 
 learning a frame-level audio-to-label mapping
 without considering the characteristics of chord label sequences
 and the generative process of music.
In addition,
 the model needs to be trained in a fully-supervised manner.
Manual chord annotation is a 
 labor-intensive task,
 and the performance of ACE
 heavily depends on the amount, diversity 
 and quality
 of annotated music signals\cite{Humphrey2015}
 because of the nature of supervised learning.

Given that the performance of ACE could be improved 
 by integrating the complementary generative and discriminative approaches,
 we propose a semi-supervised neural chord estimator
 based on a variational autoencoder (VAE)\cite{kingma2014vae} (Fig.~\ref{fig:overview}).
Because chroma vectors can be represented more precisely 
 by considering their continuous latent features 
 (deviations from basic chroma patterns)
 in addition to the underlying discrete chord labels,
 we formulate a DNN-based \textit{generative} model $p_\theta$ 
 that represents the generative process of 
 a sequence of chroma vectors
 from that of chord labels and that of latent features.
To complete the Bayesian formulation,
 we introduce prior distributions on the latent variables.
Specifically, the chord labels are assumed to follow 
 a first-order Markov model favoring self transitions,
 and the latent features, which abstractly
 represent the fine structure of the chroma vectors, 
 are assumed to follow a standard Gaussian distribution.
In the framework of amortized variational inference (AVI)~\cite{Gershman2014amortized},
 DNN-based \textit{discriminative} models $q_\alpha$ and $q_\beta$ 
 are then introduced as variational posterior distributions
 to approximate the posterior distribution of the latent variables 
 from an observed chroma sequence.
The generative and discriminative models 
 can be trained jointly in a semi-supervised manner 
 by using music signals with and without chord annotations.
 
The main contribution of this paper is 
 to draw the potential of the powerful deep discriminative model for ACE,
 by integrating it into the principled statistical inference formalism 
 of the generative approach. 
This is the first attempt to use a VAE for semi-supervised ACE.
A key feature of our VAE 
 is that a Markov model is used to
 define the prior probability
 of chord label sequences,
 which works as a chord language model
 and prevents frequent frame-level transitions of chord labels 
 in the joint training of the generative and discriminative models 
 for given chroma vectors.
We experimentally show the 
 effectiveness of this regularization on semi-supervised learning.

The remainder of this paper is organized as follows.
Section \ref{sec:related_work} reviews related work 
 on the generative and discriminative approaches to ACE.
Section~\ref{sec:proposed_method} describes the proposed ACE method
 based on the semi-supervised VAE 
 with the Markov prior on latent chord labels.
Section~\ref{sec:evaluation} reports the experimental results
 and discusses the effectiveness and limitations of the proposed techniques.
Section~\ref{sec:conclusion} summarizes the paper.

\section{Related Work}
\label{sec:related_work}

This section reviews related work on ACE
 based on generative and discriminative approach,
 and machine-learning strategies for integrating these approaches.
 
\subsection{Generative Approach}

In the generative approach,
 a probabilistic generative model
 $p(\mX,\mS) = p(\mX|\mS) p(\mS)$ has often been formulated as an HMM\cite{Rabiner1989},
 where $\mX$ and $\mS$ denote a sequence of acoustic features 
 (e.g., chroma vectors \cite{fujishima1999})
 and a sequence of chord labels  \cite{Sheh2003}, respectively.
The emission probabilities of $\mX$ for each chord, $p(\mX|\mS)$,
 and the chord transition probabilities, $p(\mS)$,
 are typically given by a Gaussian mixture model (GMM)
 and a first- or higher-order Markov model (n-gram model), respectively.
Given $\mX$ as observed data, 
 the optimal chord label sequence maximizing the posterior probability $p(\mS|\mX)$
 can be estimated efficiently by using the Viterbi algorithm.

Many HMM-based methods have been proposed for ACE.
Lee and Slaney \cite{Lee2008}, for example, 
 proposed joint training of multiple HMMs
 corresponding to different keys. %
Another research direction
 is to simultaneously deal with multiple kinds of musical elements 
 in addition to chord labels
 by explicitly considering the relationships of those elements.
Mauch and Dixon \cite{Mauch2010} 
 proposed a dynamic Bayesian network (DBN)
 that represents the hierarchical structure
 over metrical positions, musical keys, bass notes, and chords, 
 and formulates the generative process of treble and bass chroma vectors.
Similarly, Ni \etal \cite{Ni2012} proposed 
 a harmony progression analyzer (HPA) 
 based on a DBN whose latent states represent 
 chords, inversions, and musical keys.

Some studies have focused on the temporal characteristics 
 of frame-level chord labels. 
To represent repetitions of chord patterns, for example,
 multi-order HMMs\cite{khadkevich2009}
 and duration-explicit HMMs\cite{Chen2012} have been proposed,
 where the results of ACE were found to be insensitive
 to the frame-level transition probabilities of chord labels \cite{Chen2012}.
Although such a frame-level language model
 is incapable of learning typical chord progressions at the symbol level, 
 it is still effective for encouraging the continuity of chord labels at the frame level.

\subsection{Discriminative Approach}

Recently, DNNs have intensively been used
 as a powerful discriminative model
 for directly estimating the posterior probability $p(\mS|\mX)$.
Humphrey and Bello \cite{humphrey2012rethinking}, for example,
 used a convolutional neural network (CNN)
 for learning an effective representation
 of raw audio spectrograms.
A number of ACE methods using DNN classifiers 
 have then been proposed for using low-level audio representations
 such as constant-Q transform (CQT) spectrograms\cite{schorkhuber2010}.
In general, 
 these DNN-based methods outperform
 HMM-based generative methods\cite{Korzeniowski2016}.

Because typical DNN-based methods estimate 
 the posterior probabilities of chord labels at the frame level,
 some smoothing technique is often used 
 for estimating temporally-coherent chord labels.
An HMM \cite{wu2019eusipco} 
 or a conditional random field (CRF)~\cite{Korzeniowski2016}, for example,
 can be used for estimating the optimal path of chord labels
 from the estimated posterior probabilities.
Recurrent neural networks (RNNs) have recently been used
 as a language model that represents the long-term dependency of chord labels
 \cite{boulanger2013audio,Sigtia2015}.
Note that RNN-based models are still
 incapable of learning the symbol-level syntactic structure 
 from frame-level chord sequences\cite{korzeniowski2017on}.
To solve this problem,
 Korzeniowski and Widmer \cite{Korzeniowski2018} 
 formulated a symbol-level chord transition model with a duration distribution.
Chen and Su \cite{Chen2019Harmony} proposed an extension of the transformer model \cite{attention2017}
 for joint chord segmentation and labeling in a context of multi-task learning.
 
Even the state-of-the-art neural chord estimators 
 have difficulty in identifying infrequent chord types
 because the numbers of occurrences of chord types are highly imbalanced.
In fact, the majority of chords is occupied by major and minor triads.
Another difficulty lies in distinguishing chords 
 that commonly include several pitch classes as chord notes (\eg, C9 and Csus2).
When the chord vocabulary gets larger, 
 \ie, the chord classes are defined at a finer level,
 the subjectivity of annotators 
 leads to the inconsistency of chord annotations~\cite{Hendrik2019}.
This would make infrequent chord types harder to learn.
This problem can be mitigated
 by reflecting the musical knowledge
 about the constituent tones and taxonomy of chord labels \cite{Mcfee2017,CarsaultNE18,jiang2019}
 into objective functions
 and/or by using an even-chance training scheme \cite{Deng2017}.
However, dealing with a large chord vocabulary is still an open problem.

\subsection{Unified Generative and Discriminative Approach}

Integration of deep generative and discriminative models
 has actively been explored 
 in the context of unsupervised or semi-supervised learning.
One of the most popular strategies is
 to use a variational autoencoder (VAE)~\cite{kingma2014vae}
 that jointly optimizes a deep generative model $p(\mX|\mS)$
 and a deep recognition model $q(\mS|\mX)$
 that approximates $p(\mS|\mX)$
 such that the lower bound of the marginal likelihood $p(\mX)$ is maximized.
An extension, known as conditional VAE\cite{kingma2014semi},
 can be used for semi-supervised learning of latent representations
 disentangled from given labels (conditions)
 \cite{Maale2016Auxiliary}.
In the field of Automatic Speech Recognition (ASR),
 some studies tried to jointly train 
 a speech-to-text model with a text-to-speech model
 to improve the performance of ASR
 by using both annotated and non-annotated speech signals \cite{Hori2018}.
 
In this paper, we use a VAE 
 with two different latent variables
 corresponding to chord labels (\textit{categorical} variables)
 and latent features (\textit{continuous} variables).
This model is similar to JointVAE \cite{dupont2018}
 in a sense that both discrete and continuous representations are learned jointly.
In our model, 
 a Markov prior favoring self transitions is put on chord labels
 for regularizing the recognition model
 in the VAE training.
In conventional studies, in contrast,  
 the Markov prior is used only 
 for smoothing chord labels estimated by a recognition model.
To our knowledge, our work is the first attempt
 to integrate the generative and discriminative processes 
 between chord labels and acoustic features
 into a unified jointly-trainable DNN.

\section{Proposed Method}
\label{sec:proposed_method}

\begin{figure}[t]
 \centering
 \centerline{\includegraphics[width=.75\columnwidth]{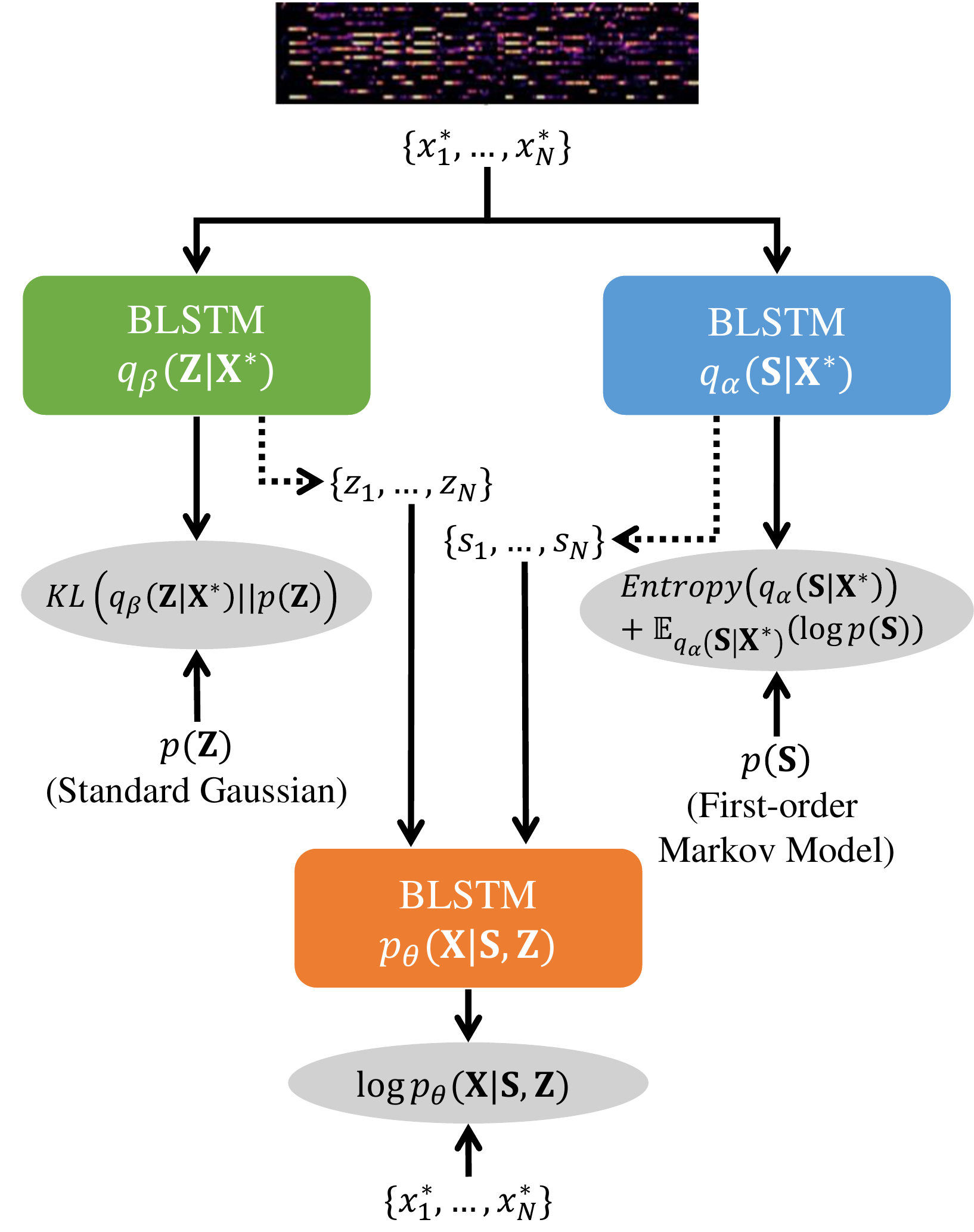}}
 \vspace{-2mm}
 \caption{The computation flow of the unsupervised learning. 
 The gray areas correspond to the four terms of the cost function $\mathcal{L}_\mX$ 
 given by (\ref{eq:objective_unsupervised}).}
 \label{fig:calcflow_unsupervised}
 \vspace{-2mm}
\end{figure}

This section describes 
 the proposed unified generative and discriminative approach to ACE
 (Fig.~\ref{fig:calcflow_unsupervised}).
To tackle the frame-level ACE (Section~\ref{sec:problem_specification}),
 we formulate a probabilistic model $p_\theta$ representing 
 the generative process of chroma vectors from chord labels and latent features
 (Section~\ref{sec:generative_model}),
 and then introduce neural statistical estimators $q_\alpha$ and $q_\beta$ 
 that respectively infers
 chord labels and latent features from chroma vectors (Section~\ref{sec:inference_models}).
These three models are jointly trained
 in the framework of amortized variational inference (AVI)\cite{Gershman2014amortized}.
In theory, $q_\alpha$ can be trained in an unsupervised manner
 only from chroma vectors without referring to their chord labels
 (Section~\ref{sec:unsupervised_training}).
In practice, $q_\alpha$ is trained
 in a supervised or semi-supervised manner
 by using paired data 
 (Sections~\ref{sec:supervised_training}).

\subsection{Problem Specification}
\label{sec:problem_specification}

For simplified explanation,
 suppose that we have one musical piece.
Let $\mX=\{\vx_1,\cdots,\vx_N\}$ be 
 a sequence of chroma vectors extracted from the music signal,
 where $\vx_n \in \{0, 1\}^{D}$ is a $D$-dimensional binary vector
 representing the activations of $D$ pitch classes at frame $n$,
 and $N$ is the number of frames.
In this paper, $\vx_n$ is a 3-channel chroma vector,
 each channel representing the
 lower, middle, and higher pitch ranges respectively ($D = 12 \times 3 = 36$),
 as defined in \cite{Wu2019}.
Because the true pitch activations are not given in reality,
 $\vx_n$ is obtained 
 by using a neural multi-pitch estimator \cite{Wu2019}
 that estimates frame-wise pitch-class distributions
 with a DNN trained from 6,000 pieces
 (audio signals synthesized from MIDI files)\footnote{
 https://github.com/Xiao-Ming/ChordRecognitionMIDITrainedExtractor}.
We thus relax the binary constraint and assume $\vx_n \in [0, 1]^{D}$.

Let $\mS=\{\vs_1,\cdots,\vs_N\}$ be
 a sequence of chord labels corresponding to $\mX$,
 where $\vs_n\in\{0,1\}^K$ is a categorical variable
 ($K$-dimensional one-hot vector)
 indicating the chord label of frame $n$.
In this paper, the chord vocabulary consists of 
 all possible combinations of twelve root notes
 with six types of triad chords
 (with shorthands \textit{maj}, \textit{min}, \textit{dim}, \textit{aug}, 
 \textit{sus2}, and \textit{sus4})
 and two types of power chords (with shorthands \textit{1} and \textit{5}),
 and a no-chord label (with shorthand \textit{N}),
 \ie, $K=12 \times 8 + 1 = 97$.
We follow Harte's notation \cite{Harte2010}, \eg, C:maj and C\#:sus4.

Let $\mZ=\{\vz_1,\cdots,\vz_N\}$ be a sequence of latent features,
 where $\vz_n \in\mathbb{R}^{L}$ is a continuous variable
 that abstractly represents 
 how $\vx_n$ is deviated from a basic chroma pattern specified by $\vs_n$
 ($L=64$).
\ie, not only $\vs_n$ but also $\vz_n$ are needed to fully describe $\vx_n$.
If a region of C:maj includes some passing notes with different volumes, 
 for example, the chroma vectors are deviated from a basic pattern
 that takes one at the dimensions corresponding to C, E, and G.

Our goal is to train a classification model $p(\mS|\mX)$
 and use it for estimating chord labels behind unseen music signals.
In an unsupervised condition,
 the classification model should be trained
 from $\mX$ without using the ground-truth data of $\mS$.
In a supervised condition,
 the classification model can be trained
 by using paired data of $\mX$ and $\mS$.
In a semi-supervised condition,
 some part of $\mS$ is given as ground-truth data.
 
\subsection{Generative Model}
\label{sec:generative_model}

We formulate a probabilistic hierarchical generative model
 (joint probability distribution) of 
 chord labels $\mS$, latent features $\mZ$, and chroma vectors $\mX$ as follows:
\begin{align}
 p_{\theta,\phi}(\mX, \mS, \mZ) = p_\theta(\mX | \mS, \mZ) p_\phi(\mS) p(\mZ),
 \label{eq:p_x_s_z}
\end{align}
 where $p_{\theta}(\mX | \mS, \mZ)$ 
 is a likelihood function of $\mS$ and $\mZ$ for $\mX$,
 $p_\phi(\mS)$ is a prior distribution of $\mS$,
 $p(\mZ)$ is the prior of $\mZ$,
 and $\theta$ and $\phi$ are model parameters.
While standard HMMs used for ACE are represented as
 $p_{\theta,\phi}(\mX, \mS) = p_{\theta}(\mX | \mS) p_{\phi}(\mS)$~\cite{McVicar2014},
 we use both $\mS$ and $\mZ$ for precisely representing $\mX$,
 \ie,
 $p_\theta(\mX | \mS, \mZ)$ is a deep \textit{generative} model 
 represented as follows:
\begin{align}
 p_{\theta}(\mX | \mS, \mZ) 
 = 
 \prod_{n=1}^N \prod_{d=1}^D
 \mathrm{Bernoulli}(x_{nd} | [\bm\omega_{\theta}(\mS,\mZ)]_{nd}),
 \label{eq:p_x}
\end{align}
where $\bm\omega_\theta(\mS,\mZ)$ is the $ND$-dimensional output of 
 a DNN with parameters $\theta$
 that takes $\mS$ and $\mZ$ as input
 and the notation $[\mA]_{ij}$ indicates the $ij$-th element of $\mA$.
Although $x_{nd}$ should take a binary value in theory,
 we allow it to take a real value between $0$ and $1$
 from a practical point of view.

If we have no prior knowledge on chord labels $\mS$,
 a natural choice of $p_\phi(\mS)$ is a uniform distribution as follows:
\begin{align}
p_\phi(\mS) = \prod_{n=1}^N 
\mathrm{Categorical}
\left(\vs_n | \textstyle\left[\frac{1}{K},\cdots,\frac{1}{K}\right]\right),
\label{eq:p_s_uni}
\end{align}

Since chord labels $\mS$ have temporal continuity, 
 \ie, change infrequently at the frame level,
 we follow the common assumption on chord labels~\cite{Sheh2003}
 and use a first-order Markov model favoring self-transitions of chord labels
 as the chord label prior $p_\phi(\mS)$ as follows:
\begin{align}
 p_{\phi}(\mS)
 &= p(\vs_1) \prod_{n=2}^N p(\vs_n | \vs_{n-1}) 
 \nonumber\\
 &= 
 \prod_{k=1}^K 
 \phi_{k}^{s_{1k}}
 \prod_{n=2}^N \prod_{k'=1}^K \prod_{k=1}^K 
 \phi_{k'k}^{s_{n-1,k'} s_{nk}},
 \label{eq:p_s}
\end{align}
where $\phi_{k}$ is the initial probability of chord $k$
 and $\phi_{k'k}$ is the transition probability
 from chord $k'$ to chord $k$.
Since $\phi_{k}$ and $\phi_{k'k}$
represent the probabilities,
$\sum_{k=1}^K{\phi_{k}=1}$, $\sum_{k=1}^K{\phi_{k'k}=1}$.
In the ACE researches based on HMMs, the transition probability matrices
 typically have high self-transition probability,
 which reflects the continuity of chord label sequences\cite{Chen2012}.
In this paper, we set $\phi_{kk}=0.9$.
The effects of (\ref{eq:p_s_uni}) an (\ref{eq:p_s})
 are compared in Section~\ref{sec:effect_markov_prior}.

Since we have no strong belief
 about the abstract latent features $\mZ$,
 the prior $p(\mZ)$ 
 is set to a standard Gaussian distribution as follows:
\begin{align}
 p(\mZ)
 =
 \prod_{n=1}^N \mathcal{N}(\vz_n | \mathbf{0}_L, \mathbf{I}_L),
 \label{eq:p_z}
\end{align}
where $\mathbf{0}_L$ is the all-zero vector of size $L$
 and $\mathbf{I}_L$ is the identity matrix of size $L\times L$.

\subsection{Classification and Recognition Models}
\label{sec:inference_models}

Given chroma vectors $\mX$ as observed data,
 we aim to infer the chord labels $\mS$
 and the latent features $\mZ$ from $\mX$
 and estimate the model parameters $\theta$ and $\phi$
 in the maximum-likelihood framework.
Because the DNN-based formulation makes
 the posterior distribution 
 $p_{\theta,\phi}(\mS,\mZ|\mX) \propto p_{\theta,\phi}(\mX,\mS,\mZ)$ 
 analytically intractable,
 we compute it approximately with an AVI technique.
More specifically, 
 we introduce a sufficiently-expressive variational distribution 
 $q_{\alpha,\beta}(\mS,\mZ|\mX)$
 parametrized by $\alpha$ and $\beta$
 and optimize it such that the Kullback-Leibler (KL) divergence
 from $q_{\alpha,\beta}(\mS,\mZ|\mX)$ to $p_{\theta,\phi}(\mS,\mZ|\mX)$ is minimized.
Considering that 
 both $\mS$ and $\mZ$ make an effect 
 on the generative model $p_{\theta}(\mX|\mS,\mZ)$,
 they are assumed to be conditionally independent 
 in the inference model $q_{\alpha,\beta}(\mS,\mZ|\mX)$
 as follows:
\begin{align}
    q_{\alpha,\beta}(\mS,\mZ|\mX) = q_\alpha(\mS|\mXalt)q_\beta(\mZ|\mXalt),
    \label{eq:q_s_z}    
\end{align}
where $q_\alpha(\mS|\mXalt)$ and $q_\beta(\mZ|\mXalt)$ are
\textit{classification} and \textit{recognition} models
that infer $\mS$ and $\mZ$, respectively.
In our study,
 (\ref{eq:q_s_z}) was found to work better 
 than $q_{\alpha,\beta}(\mS,\mZ|\mX) = q_\alpha(\mS|\mXalt)q_\beta(\mZ|\mXalt,\mS)$
 respecting the chain rule of probability as in \cite{kingma2014semi}.
In this paper, these models are implemented
with DNNs parameterized by $\alpha$ and $\beta$ as follows:
\begin{align}
q_\alpha(\mS|\mXalt)
&=
\prod_{n=1}^N \mathrm{Categorical}(\vs_n | [\bm\pi_\alpha(\mXalt)]_n),
\label{eq:q_s}
\\
q_\beta(\mZ|\mXalt)
&= \prod_{n=1}^N
\mathcal{N}(\vz_n | [\bm\mu_\beta(\mXalt)]_n, [\bm\sigma^2_\beta(\mXalt)]_n),
\label{eq:q_z}
\end{align}
where $\bm\pi_\alpha(\mX)$ is the $NK$-dimensional output 
 of the DNN with parameters $\alpha$,
 and $\bm\mu_\beta(\mXalt)$ and $\bm\sigma^2_\beta(\mXalt)$ are 
 the $NL$-dimensional outputs of the DNN with parameters $\beta$.
Similar to the deep generative model,
 the outputs of the DNNs represent the parameters of 
 probabilistic distributions. 
 
\subsection{Unsupervised Learning with Non-Annotated Data}
\label{sec:unsupervised_training}

Instead of directly maximizing the log-marginal likelihood $\log p_{\theta,\phi}(\mX)$
 with respect to the model parameters $\theta$ and $\phi$,
 we maximize its variational lower bound $\mathcal{L}_\mX(\theta,\phi,\alpha,\beta)$ 
 derived by introducing $q_{\alpha,\beta}(\mS,\mZ|\mX)$ as follows:
\begin{align}
 \log p_{\theta,\phi}(\mX) 
　&= 
\log \iint p_{\theta,\phi}(\mX, \mS, \mZ) d\mS d\mZ
\nonumber\\
 &=
 \log \int\!\!\!\!\int\! 
 \frac{q_{\alpha,\beta}(\mS,\mZ|\mX)}{q_{\alpha,\beta}(\mS,\mZ|\mX)}  p_{\theta,\phi}(\mX, \mS, \mZ) d\mS d\mZ
\nonumber\\
 &\ge
 \int\!\!\!\!\int\! 
 q_{\alpha,\beta}(\mS,\mZ|\mX) \log \frac{p_{\theta,\phi}(\mX, \mS, \mZ)}{q_{\alpha,\beta}(\mS,\mZ|\mX)} d\mS d\mZ
\nonumber\\
 &\defeqn \mathcal{L}_\mX(\theta,\phi,\alpha,\beta),
 \label{eq:lower_bound_unsup}
\end{align}
where the equality holds, 
\ie, $\mathcal{L}_\mX(\theta,\phi,\alpha,\beta)$ is maximized,
if and only if $q_{\alpha,\beta}(\mS,\mZ|\mX)=p_{\theta,\phi}(\mS,\mZ|\mX)$.
Note that this condition cannot be satisfied
 because $p_{\theta,\phi}(\mS,\mZ|\mX)$ is hard to compute.
Because the gap between $\log p_{\theta,\phi}(\mX)$ 
 and $\mathcal{L}_\mX(\theta,\phi,\alpha,\beta)$ in (\ref{eq:lower_bound_unsup})
 is equal to the KL divergence
 from $q_{\alpha,\beta}(\mS,\mZ|\mX)$ to $p_{\theta,\phi}(\mS,\mZ|\mX)$,
 the minimization of the KL divergence 
 is equivalent to the maximization of $\mathcal{L}_\mX(\theta,\phi,\alpha,\beta)$.

To approximately compute $\mathcal{L}_\mX(\theta,\phi,\alpha,\beta)$,
 we use Monte Carlo integration as follows:
    \begin{align}
         &\mathcal{L}_\mX(\theta,\phi,\alpha,\beta)
         \nonumber\\
         &=\mathbb{E}_{q_\alpha(\mS|\mXalt)q_\beta(\mZ|\mXalt)}
         [\log p_\theta(\mX|\mS,\mZ)]
         \nonumber\\
         &\quad+
         \mathbb{E}_{q_\alpha(\mS|\mXalt)q_\beta(\mZ|\mXalt)}
         [\log p(\mZ)-\log q_\beta(\mZ|\mXalt)]
         \nonumber\\
         &\quad+
         \mathbb{E}_{q_\alpha(\mS|\mXalt)}
         [\log p_\phi(\mS)-\log q_\alpha(\mS|\mXalt)]
         \nonumber\\
         &\approx \frac{1}{I}\sum_{i=1}^I 
         \log p_\theta(\mX|\mS_i,\mZ_i)-\mathrm{KL}(q_\beta(\mZ|\mXalt)||p(\mZ))
         \nonumber\\
         &\quad+
         \mathrm{Entropy}[q_\alpha(\mS|\mXalt)] 
         +
         \mathbb{E}_{q_\alpha(\mS|\mXalt)}[\log p_\phi(\mS)],
    \label{eq:objective_unsupervised}
    \end{align}
where $\{\mS_i,\mZ_i\}_{i=1}^I$ are $I$ samples drawn from $q_\alpha(\mS|\mXalt)q_\beta(\mZ|\mXalt)$.
Following the typical VAE implementation,
 we set $I=1$ and hence the index $i$ can be omitted.
To make (\ref{eq:objective_unsupervised}) partially differentiable
 with respect to the model parameters $\theta$, $\phi$, $\alpha$, and $\beta$,
 we use reparametrization tricks~\cite{kingma2014vae,Eric2017}
 for deterministically representing the categorical variables $\mS$ 
 and the Gaussian variables $\mZ$ as follows:
\begin{align}
\bm\epsilon^\vs_n 
&\sim 
\mathrm{Gumbel}(\mathbf{0}_K,\mathbf{1}_K), 
\label{eq:e_s_n}
\\
\vs_n 
&= 
\mathrm{softmax}
((\log [\bm\pi_\alpha(\mXalt)]_n + \bm\epsilon^\vs_n)/\tau),
\label{eq:s_n_rt}
\\
\bm\epsilon^\vz_n 
&\sim 
\mathcal{N}(\mathbf{0}_L,\mathbf{I}_L), 
\label{eq:e_z_n}
\\
\vz_n 
&= 
[\bm\mu_\beta(\mXalt)]_n 
+ \bm\epsilon^\vz_n \odot [\bm\sigma_\beta(\mXalt)]_n,
\label{eq:z_n_rt}
\end{align}
where (\ref{eq:e_s_n}) indicates the standard Gumbel distribution,
 $\mathbf{1}_K$ is the all-one vector of size $K$,
 $\odot$ means the element-wise product,
 and $\tau > 0$ is a temperature parameter
 that controls the uniformity of $\vs_n$
 ($\tau=0.1$ in this paper).

We can now compute the four terms of (\ref{eq:objective_unsupervised}) 
 evaluating the fitness of $\mS$ and $\mZ$:
 a reconstruction term indicating the likelihood of $\mS$ and $\mZ$ for $\mX$
 and three regularization terms
 making the posterior $q_\beta(\mZ|\mXalt)$ close to the prior $p(\mZ)$,
 increasing the entropy of $q_\alpha(\mS|\mXalt)$,
 and making $\mS$ temporally coherent, respectively
 (Fig.~\ref{fig:calcflow_unsupervised}).
More specifically,
 the first term is given by (\ref{eq:p_x}) 
 and the second and third terms are given by 
\begin{align}
 &\mathrm{KL}(q_\beta(\mZ|\mXalt)||p(\mZ))
 \nonumber\\
 &
 = \!
 \sum_{n=1}^N\sum_{k=1}^K
 \!
 \left(\frac{[\bm\sigma_\beta(\mX)]_{nk}^2 \! 
 + \! [\bm\mu_\beta(\mX)]_{nk}^2 \! - \! 1}{2} 
 -\log([\bm\sigma_\beta(\mX)]_{nk})\!
 \right)\!,
 \\
 &\mathrm{Entropy}[q_\alpha(\mS|\mXalt)] 
 \nonumber\\
 &= 
 -\sum_{n=1}^N\sum_{k=1}^K[\pi_\alpha(\mX)]_{nk}
 \log([\pi_\alpha(\mX)]_{nk}).
 \label{eqn:exp_entropy_s}
\end{align}

The last term $\mathbb{E}_{q_\alpha(\mS|\mXalt)}[\log p_\theta(\mS)]$
 can be calculated analytically.
If $p_\theta(\mS)$ is the uniform distribution given by (\ref{eq:p_s_uni}),
 we have
\begin{align}
\mathbb{E}_{q_\alpha(\mS|\mXalt)}[\log p(\mS)]
=
- \sum_{n=1}^N \sum_{k=1}^K [\bm\pi_\alpha(\mXalt)]_{nk} \log K.
\end{align}
If $p_\theta(\mS)$ is the Markov model given by (\ref{eq:p_s}),
 we use a dynamic programming technique
 similar to the forward algorithm of the HMM.
Let $\gamma(\vs_n)=\mathbb{E}_{q_\alpha(\mS|\mX)}[\log p(\vs_{1:n})]$
 be a forward message at frame $n$,
 which can be calculated recursively as follows:
 \begin{align}
&\gamma(\vs_1) = 
\log p(\vs_1),
\\
&\gamma(\vs_n) = 
\sum_{\vs_{n-1}}q_\alpha(\vs_{n-1}|\mXalt)
\bigl(\gamma(\vs_{n-1})+\log p_\phi(\vs_n|\vs_{n-1})\bigr), 
\\
&\mathbb{E}_{q_\alpha(\mS|\mXalt)}[\log p_\phi(\mS)] 
= \sum_{\vs_N} q_\alpha(\vs_N|\mXalt)\gamma(\vs_N).
 \label{eqn:exp_prior_s}
\end{align}

We now have a VAE (Fig.~\ref{fig:calcflow_unsupervised}) that consists of:
\begin{itemize}
\item
 the deep classification model 
 $q_\alpha(\mS|\mXalt)$ given by~(\ref{eq:q_s})
 with the reparametrization trick given by~(\ref{eq:e_s_n}) and (\ref{eq:s_n_rt}),
 \item
 the deep recognition model 
 $q_\beta(\mZ|\mXalt)$ given by~(\ref{eq:q_z})
 with the reparametrization trick given by~(\ref{eq:e_z_n}) and (\ref{eq:z_n_rt}), and
 \item
 the deep generative model $p_{\theta}(\mX | \mS, \mZ)$ 
 given by~(\ref{eq:p_x})
 with the prior distributions $p_{\phi}(\mS)$ and $p(\mZ)$ 
 given by (\ref{eq:p_s}) and (\ref{eq:p_z}), respectively.
\end{itemize}
In the unsupervised condition,
 all models are jointly optimized in the framework of the VAE
 by using a variant of stochastic gradient descent
 such that (\ref{eq:objective_unsupervised}) is maximized
 with respect to $\theta$, $\alpha$, and $\beta$.

\subsection{Supervised Learning with Annotated Data}
\label{sec:supervised_training}

Under the supervised condition
 that chroma vectors $\mX$ and the corresponding chord labels $\mS$ 
 are given as observed data,
 we aim to maximize a variational lower bound $\mathcal{L}_{\mXalt,\mS}(\theta,\beta)$ 
 of the log-likelihood $\log p_\theta(\mS|\mX)$,
 which can be derived in a way similar to (\ref{eq:lower_bound_unsup}) as follows:
\begin{align}
 \log p_\theta(\mX|\mS)
 &=
 \log\int\!
 \frac{q_\beta(\mZ|\mXalt)}{q_\beta(\mZ|\mXalt)}
 p_\theta(\mX|\mZ,\mS)d\mZ
 \nonumber\\
 &\ge 
 \int q_\beta(\mZ|\mXalt)
 \log\frac{p_\theta(\mX|\mZ,\mS)}{q_\beta(\mZ|\mXalt)}d\mZ
 \nonumber\\    
 &\defeqn
 \mathcal{L}_{\mXalt,\mS}(\theta,\beta).
 \label{eq:lower_bound_2}
\end{align}
Using  Monte Carlo integration, 
 $\mathcal{L}_{\mXalt,\mS}(\theta,\beta)$
 can be approximately computed as follows:
\begin{align}
 &\mathcal{L}_{\mX,\mS}(\theta,\beta)
 \nonumber\\
 &\defeqn
 \mathbb{E}_{q_\beta(\mZ|\mXalt)}(\log p_\theta(\mX|\mS,\mZ)+\log p(\mZ)
 - \log　q_\beta(\mZ|\mXalt))
 \nonumber\\
 &\approx  
 \frac{1}{I}
 \sum_{i=1}^I 
 \log p_\theta(\mX | \mS_i, \mZ_i)
+\mathrm{KL}(q_\beta(\mZ|\mXalt) \| p(\mZ)),
\label{eq:deviation_supv}
 \end{align}
where $\{\mZ_i\}_{i=1}^I$ are $I$ samples
 drawn from $q_\beta(\mZ|\mX,\mS)$
 using the reparametrization trick ($I=1$ in this paper).    
Because the chord estimator $q_\alpha(\mS|\mXalt)$,
 which plays a central role in ACE,
 does not appear in (\ref{eq:deviation_supv}),
 $q_\alpha(\mS|\mXalt)$ cannot be trained
 only by maximizing (\ref{eq:deviation_supv}).
As suggested in the semi-supervised learning of a VAE \cite{kingma2014semi},
 one could thus define an alternative objective function 
 $\mathcal{L}_{\mX,\mS}(\theta,\alpha,\beta)$ including $\alpha$
 by adding a classification performance term to (\ref{eq:deviation_supv}) as follows:
\begin{align}
 \mathcal{L}_{\mX,\mS}(\theta,\alpha,\beta)
 \defeqn
 \mathcal{L}_{\mX,\mS}(\theta,\beta) + \log q_\alpha(\mS|\mXalt).
 \label{eq:objective_supervised_org}
\end{align} 
A problem of this approach, however,
 is that the regularization terms 
 (\ref{eqn:exp_entropy_s}) and (\ref{eqn:exp_prior_s})
 enhancing the entropy of $q_\alpha(\mS|\mXalt)$ (preventing the overfitting)
 and smoothing the output of $q_\alpha(\mS|\mXalt)$, respectively,
 are not taken into account.

To solve this problem,
 we propose a new objective function $\mathcal{L}_{\mX,\mS}(\theta,\phi,\alpha,\beta)$ 
 by summing (\ref{eq:objective_unsupervised}) and (\ref{eq:objective_supervised_org}) as follows:
\begin{align}
 \mathcal{L}_{\mX,\mS}(\theta,\phi,\alpha,\beta)
 =
 {\mathcal{L}_\mX(\theta,\phi,\alpha,\beta)+
 \mathcal{L}_{\mX,\mS}(\theta,\alpha,\beta)}.
 \label{eq:objective_supervised}
\end{align}
In this function,
 the chroma vectors $\mX$ with the annotations $\mS$ 
 are used twice as unsupervised and supervised training data
 as if they were \textit{not} annotated (first term) and as they are (second term), respectively.

\subsection{Semi-supervised Learning}
\label{sec:semi-supervised_training}

Under the semi-supervised condition
 that partially-annotated chroma vectors are available,
 we define an objective function
 by summing the objective functions
 (\ref{eq:objective_unsupervised}) and (\ref{eq:objective_supervised}) 
 corresponding to the unsupervised and supervised conditions, respectively, as follows:
\begin{align}
 &\mathcal{L}'_{\mX,\mS}(\theta,\phi,\alpha,\beta)
 \nonumber\\
 &\defeqn
 \!
 \sum_{\mbox{\scriptsize{$\mX$ w/o $\mS$}}}
 \!\!
 \mathcal{L}_{\mX}(\theta,\phi,\alpha,\beta) 
 + 
 \!
 \sum_{\mbox{\scriptsize{$\mX$ with $\mS$}}}
 \!
 \mathcal{L}_{\mX,\mS}(\theta,\phi,\alpha,\beta),
 \nonumber\\
 &=
 \ \
 \sum_{\mbox{\scriptsize{$\mX$}}}
 \mathcal{L}_{\mX}(\theta,\phi,\alpha,\beta) 
 + 
 \!
 \sum_{\mbox{\scriptsize{$\mX$ with $\mS$}}}
 \!
 \mathcal{L}_{\mX,\mS}(\theta,\alpha,\beta).
 \label{eq:objective_semi-supervised}
\end{align}
Note that $q_\alpha(\mS|\mXalt)$ can always be regularized 
 by $p_\theta(\mX|\mS,\mZ)$ and $p_\phi(\mS)$
 regardless of the availability of annotations.

\subsection{Training and Prediction}

The model parameters $\theta$, $\phi$, $\alpha$, and $\beta$ can be optimized 
 with a stochastic gradient descent method (\eg, Adam~\cite{kingma2015})
 such that the objective function (\ref{eq:objective_unsupervised}),
 (\ref{eq:objective_supervised}), or (\ref{eq:objective_semi-supervised})
 is maximized. 
Nonetheless, in this paper,
 $\phi$ is fixed (not optimized) because of its wide acceptable range
 (see Section \ref{sec:evaluation_markov}).
Under the semi-supervised condition,
 each mini-batch consists of
 non-annotated and annotated chroma vectors (50\%--50\% in this paper)
 randomly selected from the training dataset.

In the test phase,
 using the neural chord estimator $q_\alpha(\mS|\mX)$,
 a sequence of the posterior probabilities of chord labels $\mS$ 
 are calculated
 from a sequence of chroma vectors $\mX$ 
 extracted from a target music signal.
The optimal temporally-coherent path of chord labels
 is then estimated via the Viterbi algorithm 
 with the transition probabilities $\bm\phi$.

\section{Evaluation}
\label{sec:evaluation}

This section reports comparative experiments
 conducted for evaluating the effectiveness 
 of the proposed method.
Specifically, we investigate the effectiveness
 of the VAE-based regularized training,
 that of using the Markov prior on chord labels,
 and, that of using external non-annotated data.
 
\subsection{Experimental Conditions}

We here explain compared ACE methods, network configurations,
 datasets, and evaluation procedures and measures.

\subsubsection{Compared Methods}

As listed in Table~\ref{tab:conditions}, 
 we trained a chord estimator $q_\alpha(\mS|\mXalt)$
 in five different ways:
\begin{itemize}
\item \textbf{ACE-SL (baseline)}:
As in most ACE methods,
 $q_\alpha(\mS|\mXalt)$ was trained in a \textit{supervised} manner
 by minimizing the cross-entropy loss for the ground-truth labels $\mS$,
 \ie, maximizing the following objective function:
\begin{align}
\mathcal{L}_{\mX,\mS}(\alpha)
= 
\log q_\alpha(\mS|\mXalt).
\end{align}

\item \textbf{VAE-UN-SL}:
$q_\alpha(\mS|\mXalt)$ was trained in a \textit{supervised} manner
by maximizing (\ref{eq:objective_supervised}) 
with the uniform prior (\ref{eq:p_s_uni}).

\item \textbf{VAE-MR-SL}:
$q_\alpha(\mS|\mXalt)$ was trained in a \textit{supervised} manner
by maximizing (\ref{eq:objective_supervised}) 
with the Markov prior (\ref{eq:p_s}).

\item \textbf{VAE-UN-SSL}:
$q_\alpha(\mS|\mXalt)$ was trained in a \textit{semi-supervised} manner
by maximizing (\ref{eq:objective_semi-supervised}) 
with the uniform prior (\ref{eq:p_s_uni}).

\item \textbf{VAE-MR-SSL}:
$q_\alpha(\mS|\mXalt)$ was trained in a \textit{semi-supervised} manner
by maximizing (\ref{eq:objective_semi-supervised}) 
with the Markov prior (\ref{eq:p_s}).
\end{itemize}
For convenience, let `*' denote the wild-card character, 
 \eg, \textbf{VAE-*-SL} means \textbf{VAE-MR-SL} or \textbf{VAE-UN-SL}.

Comparing \textbf{ACE-SL} with \textbf{VAE-*-SL},
 we evaluated the effectiveness of the VAE architecture
 in regularizing $q_\alpha(\mS|\mXalt)$.
Comparing \textbf{VAE-*-SSL} with \textbf{VAE-*-SL},
 we evaluated the effectiveness of the semi-supervised learning.
Comparing \textbf{VAE-MR-*} with \textbf{VAE-UN-*},
 we evaluated the effectiveness of the Markov prior on $\mS$.

\begin{table}[]
\centering
\caption{Experimental conditions}
\vspace{-2mm}
\begin{tabular}{l|lll}
\Hline
 &
  \begin{tabular}[c]{@{}c@{}}Training\end{tabular} &
  \begin{tabular}[c]{@{}c@{}}Regularization\end{tabular} &
  \begin{tabular}[c]{@{}c@{}}Prior\\\end{tabular} \\ \hline
\textbf{ACE-SL (baseline)} & Supervised & NA        & NA \\
\textbf{VAE-UN-SL}        & Supervised & VAE-based & Uniform \\
\textbf{VAE-MR-SL}         & Supervised & VAE-based & Markov \\
\textbf{VAE-UN-SSL}    & Semi-supervised  & VAE-based & Uniform \\
\textbf{VAE-MR-SSL}     & Semi-supervised  & VAE-based & Markov
\\
\Hline
\end{tabular}
\label{tab:conditions}
\end{table}

\begin{table}[]
\centering
\caption{Durations of chord types in datasets used for evaluation}
\vspace{-2mm}
\begin{tabular}{c|r}
\Hline
Chord type & Duration [h] \\ \hline
maj  & 53.09            \\
min  & 16.63            \\
aug  & 0.15             \\
dim  & 0.36             \\
sus4 & 1.63             \\
sus2 & 0.25             \\
1    & 0.76             \\
5    & 0.84           \\ 
\Hline
\end{tabular}
\label{tab:shorthands}
\end{table}

\subsubsection{Network Configurations}

Each of the classifier $q_\alpha(\mS|\mX)$, the recognizer $q_\beta(\mZ|\mX)$,
 and the generator $p_\theta(\mX|\mS,\mZ)$ 
 was implemented with a three-layered BLSTM network~\cite{Graves2013BLSTM} 
 followed by layer normalization \cite{ba2016layer}.
The final layer of $q_\alpha(\mS|\mX)$ consisted of softmax functions 
 that output the frame-level posterior probabilities of $K$ chord labels.
The final layer of $q_\beta(\mZ|\mX)$ consisted of linear units
 that output $\bm\mu_\beta(\mXalt)$ and $\log \bm\sigma^2_\beta(\mXalt)$. 
The final layer of $p_\theta(\mX|\mS,\mZ)$ consisted of sigmoid functions
 that output $\bm\omega_{\theta}(\mS,\mZ)$.
Each hidden layer of the BLSTM had 256 units.
Because the architecture of $q_\alpha(\mS|\mX)$
 was similar to that of a state-of-the-art chord estimator \cite{Wu2019},
 ACE-SL was considered to be a reasonable baseline method.
 
We used Adam optimizer\cite{kingma2015},
 where the learning rate was first set to 0.001
 and then decreased exponentially by a scaling factor of 0.99 per epoch.
Gradient clipping with norm 5 was additionally applied to the optimization process.
Each mini-batch consisted of 16 sequences,
 each of which contained 645 frames (1 min).
The parameter of the Markov prior $\phi$ was fixed as stated in 
\ref{sec:generative_model},
 and $\theta$, $\alpha$ and $\beta$ were iteratively updated.
In all the methods and configurations, 
 the number of training epochs was set to 300,
 which was sufficiently large to reach convergence
 (early stopping was not used).

\subsubsection{Datasets}

We collected 1210 \textit{annotated} popular songs 
 consisting of 198 songs from Isophonics~\cite{Harte2010}, 
 100 songs from RWC-MDB-P-2001~\cite{Goto2002},
 186 songs from uspop2002~\cite{Berenzweig2004}\footnote{
 The annotations for RWC-MDB-P-2001 and uspop2002 are provided 
 by the Music and Audio Research Lab at NYU.},
 and 726 songs 
 from McGill Billboard dataset~\cite{Burgoyne2011AnEG}.
As shown in Table~\ref{tab:shorthands},
 the six types of triad chords 
 and the two types of power chords 
 are heavily imbalanced in the chord annotations,
 where most of the annotated chord labels 
 belong to the \textit{major} and \textit{minor} triads.
We also collected 700 \textit{non-annotated} popular songs
 composed by Japanese and American artists.

Each music signal sampled at 44.1kHz
 was analyzed with constant-Q transform (CQT) 
 with a shifting interval of 4096 points 
 and a frequency resolution of one semitone per bin.
The CQT spectrogram and its 1-, 2-, 3-, and 4-octave-shifted versions 
 were stacked to yield a five-layered harmonic CQT (HCQT) representation,
 which was fed to a neural multi-pitch estimator~\cite{Wu2019}
 for computing chroma vectors $\mX$.
To compensate for the imbalance of the ratios of the 12 chord roots
 in the training data,
 chroma vectors and chord labels 
 were jointly pitch-rotated 
 by a random number of semitones on each training iteration.
Pitch shifting of the multi-channel chroma vectors can be done by 
 performing vector rotation on each channel,
 and pitch shifting of chord annotations can be done by
 changing the root notes.

\subsubsection{Evaluation Procedures}

To conduct five-fold cross validation,
 we divided the 1210 annotated songs 
 into five subsets (242 songs each).
In each fold, 
 one of the subsets was kept as test data 
 and the remaining four subsets were used as training data,
 in which $I$ and $4-I$ subsets were treated 
 as annotated and non-annotated songs, respectively
 ($I \in \{1, 2, 3, 4\}$).
Not only the classifier $q_\alpha(\mS|\mX)$ 
 but also the recognizer $q_\beta(\mZ|\mX)$
 and the generator $p_\theta(\mX|\mS,\mZ)$
 were trained jointly by using all the training data (\textbf{VAE-*-SSL})
 or only the annotated data (\textbf{VAE-*-SL}).
In \textbf{ACE-SL}, in contrast, 
 only the classifier $q_\alpha(\mS|\mX)$ was trained
 by using the annotated data.

\textbf{VAE-*-SSL} was additionally tested 
 under \textit{extended} semi-supervised conditions
 that annotated 976 songs (four subsets)
 and $M$ non-annotated songs ($M \in \{250,500,700\}$)
 were used as training data in each fold.
Note that the performance was measured on the remaining 242 annotated songs,
 which did not overlap with the 700 non-annotated songs.

\subsubsection{Evaluation Measures}

The chord estimation performance (accuracy) of each method
 was measured in terms of the frame-level match rate
 between the estimated and ground-truth chord sequences.
The weighed accuracy for each song was measured
 with \textit{mir\_eval} library \cite{Raffel2014}
 in terms of the \textit{majmin} criterion
 considering only the \textit{major} and \textit{minor} triads
 plus the \textit{no-chord} label ($K=25$)
 and the \textit{triads} criterion 
 with the vocabulary defined in Section \ref{sec:problem_specification} ($K=97$).
The overall accuracy was given 
 as the average of the song-wise accuracies
 weighed by the song lengths.

\subsection{Experimental Results}
\label{sec:effect_markov_prior}

\definecolor{blue}{RGB}{0,0,255}
\definecolor{light-blue}{RGB}{31,143,254}

\definecolor{green}{RGB}{0,100,5}
\definecolor{light-green}{RGB}{124,252,26}

The overall accuracies of \textbf{ACE-SL} and \textbf{VAE-*-*}
 with respect to the numbers of annotated and non-annotated songs (denoted as $A{+}B$) used for training
 are shown in Fig.~\ref{fig:results_scores}.
To investigate the temporal continuity of $\mS$ induced by the Markov prior,
 we tested \textbf{VAE-MR-SSL} 
 with a self-transition probability $\phi_{kk} \in \{1/K, 0.3, 0.5, 0.7, 0.9\}$.
The song-wise accuracies and average chord durations at $976{+}0$ and $976{+}700$
 are compared in Fig.~\ref{fig:boxplots}.
Examples of estimated chord sequences 
 estimated by \textbf{ACE-SL}, \textbf{VAE-UN-SSL}, and \textbf{VAE-MR-SSL} with $\phi_{kk}=0.9$
 are shown in Fig.~\ref{fig:estimations}. 

\subsubsection{Evaluation of VAE-Based Regularized Training}

We confirmed the effectiveness 
 of the VAE-based regularized training of $q_\alpha(\mS|\mX)$ under all the conditions.
As shown in Fig.~\ref{fig:results_scores} and Fig.~\ref{fig:boxplots}(a),
 \textbf{VAE-*-SL} clearly outperformed \textbf{ACE-SL} by a large margin around 1 pts.
$q_\alpha(\mS|\mX)$ was regularized effectively
 by considering its entropy, the Markov or uniform prior of $\mS$,
 and the reconstruction of $\mX$ based on $\mS$ and $\mZ$.

\subsubsection{Evaluation of Semi-supervised Learning}

We confirmed the effectiveness of the semi-supervised learning of $q_\alpha(\mS|\mX)$.
Under the semi-supervised conditions
 at $244{+}732$, $488{+}488$, and $732{+}244$
 in the left half of Fig.~\ref{fig:results_scores},
 \textbf{VAE-MR-SSL} and \textbf{VAE-UN-SSL} outperformed 
 \textbf{VAE-MR-SL} and \textbf{VAE-UN-SL}, respectively,
 where the proposed \textbf{VAE-MR-SSL} worked best.
The success of the semi-supervised learning
 was attributed to the fact that 
 the musical and acoustic characteristics of the 976 songs used for training
 were consistent with those of the 244 songs used for testing
 in the five-fold cross validation with the 1210 songs.
Under the supervised condition at $976{+}0$,
 \textbf{VAE-MR-SSL} and \textbf{VAE-UN-SSL} were equivalent
 to \textbf{VAE-MR-SL} and \textbf{VAE-UN-SL}, respectively.

Under the \textit{extended} semi-supervised conditions
 at $976{+}250$, $976{+}500$, and $976{+}700$
 in the right half of Fig.~\ref{fig:results_scores},
 the performance once dropped and then barely recovered 
 according to the increase of non-annotated songs used for training.
The large difference in the musical and acoustic characteristics 
 of the annotated and non-annotated songs
 is considered to have hindered the semi-supervised learning.

\subsubsection{Evaluation of Markov Prior}
\label{sec:evaluation_markov}

\begin{figure}[t]
\centering
\includegraphics[width=.97\columnwidth,clip]{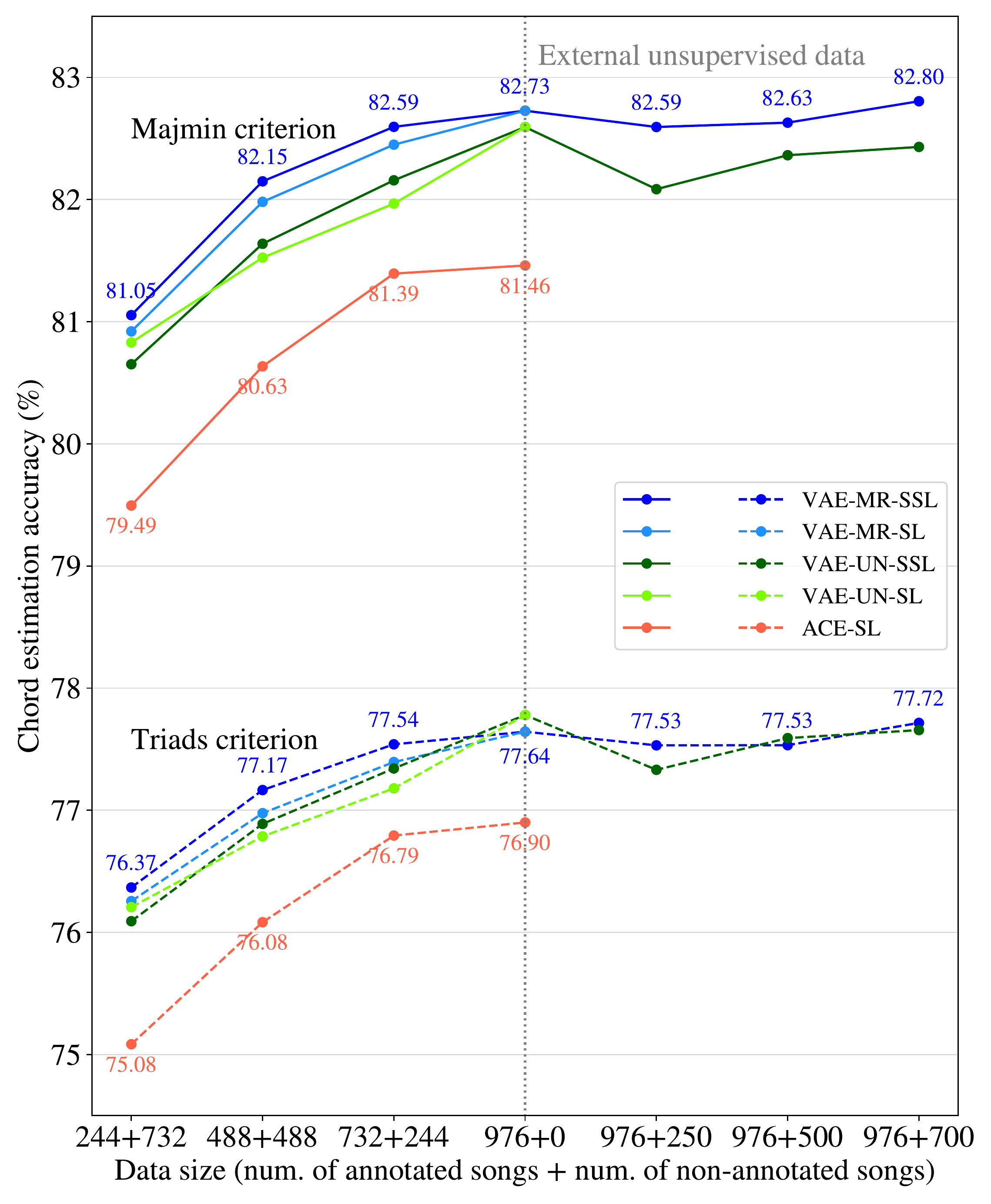}
\vspace{-3mm}
\caption{The experimental results of the five-fold cross validation 
using the 1210 annotated songs and the 700 external non-annotated songs.}
\label{fig:results_scores}
\end{figure}

We confirmed the effectiveness of the Markov prior on $\mS$
 under all the conditions.
Under the semi-supervised conditions 
 at $244{+}732$, $488{+}488$, and $732{+}244$
 and the supervised condition at $976{+}0$
 in the left half of Fig.~\ref{fig:results_scores},
 \textbf{VAE-MR-SL} and \textbf{VAE-MR-SSL}
 outperformed \textbf{VAE-UN-SL} and \textbf{VAE-UN-SSL}, respectively.
The Markov prior played a vital role 
 under the \textit{extended} semi-supervised conditions 
 at $976{+}250$, $976{+}500$, and $976{+}700$
 in the right half of Fig.~\ref{fig:results_scores}.
While \textbf{VAE-UN-SSL} at $976{+}250$
 significantly underperformed \textbf{VAE-UN-SSL} at $976{+}0$,
 \textbf{VAE-MR-SSL}
 did not experience large performance drop
 and achieved slightly better performance at $976{+}700$
 by encouraging $q_\alpha(\mS|\mX)$ 
 to yield a temporally-smooth estimate of $\mS$.

\begin{figure}[t]
\centering
\includegraphics[width=.9\columnwidth,clip]{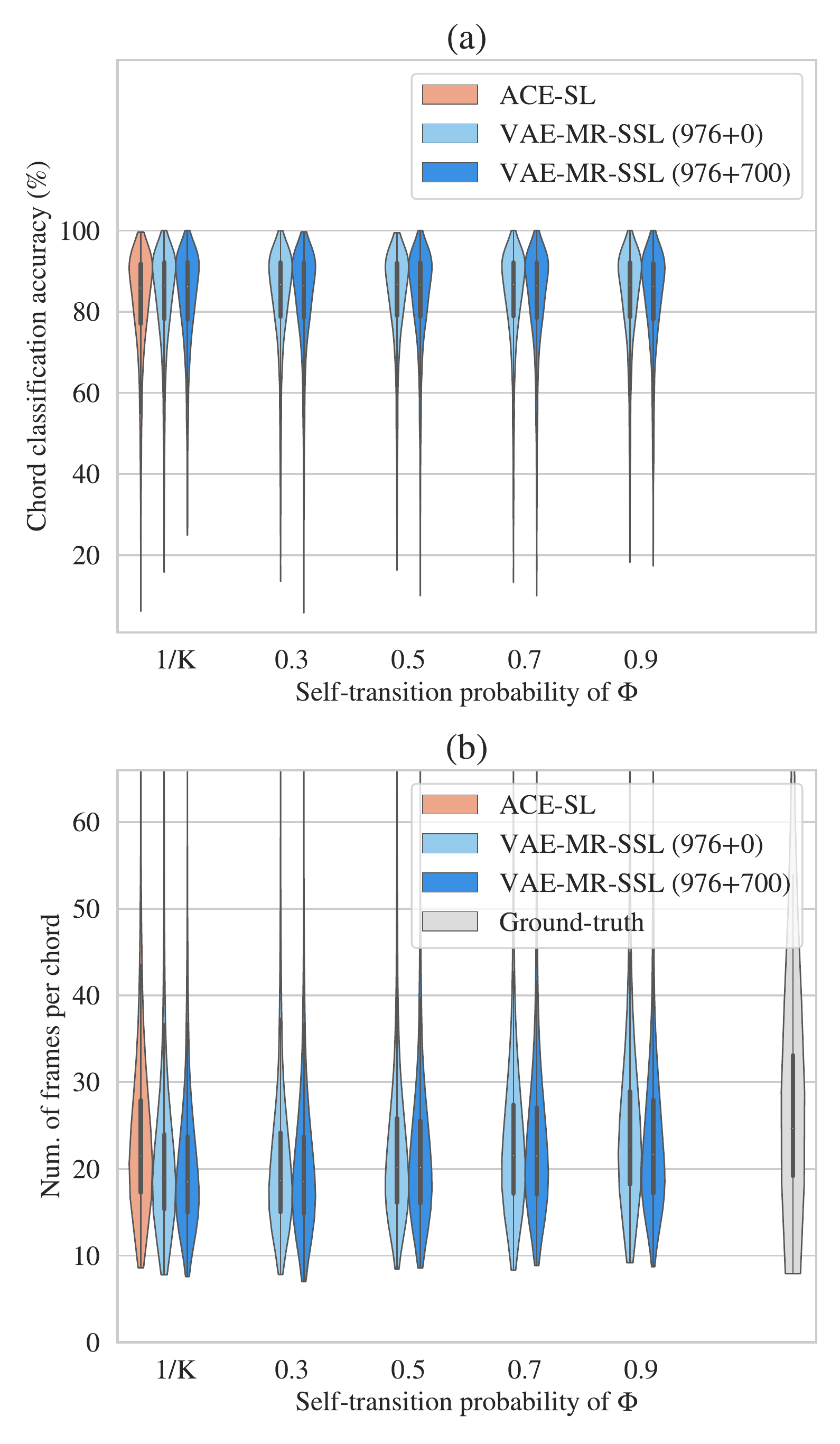}
\vspace{-3mm}
\caption{
The song-wise accuracies (in \textit{majmin} criterion) and average durations of chord labels
 estimated by \textbf{ACE-SL} and 
 \textbf{VAE-MR-SSL} with different self-transition probabilities
 ($\phi_{kk} \in \{1/K, 0.3, 0.5, 0.7, 0.9\}$),
 where \textbf{VAE-MR-SSL} with $\phi_{kk} = 1/K$ is equivalent to \textbf{VAE-UN-SSL}.
The chord durations were measured 
 before the Viterbi post-filtering.
 }
\label{fig:boxplots}
\end{figure}

Fig.~\ref{fig:boxplots}(b) shows 
 that the Markov prior mitigated the negative effect of the semi-supervised learning.
\textbf{VAE-UN-SSL} (\textbf{VAE-MR-SSL} with $\phi_{kk}=1/K$) performed better,
 but yielded finer chord segments than \textbf{ACE-SL}
 because $q_\alpha(\mS|\mX)$
 was trained under a condition that 
 chord labels were allowed to change frequently for non-annotated songs
 such that the unsupervised learning objective (\ref{eq:objective_unsupervised}) was maximized.
In contrast, \textbf{VAE-MR-SSL} with a higher self-transition probability
 yielded longer chord segments
 and the chord durations were distributed
 in a way similar to the ground-truth data.
As shown in Fig.~\ref{fig:boxplots}(a),
 the choice of the self-transition probability between $0.3$ and $0.9$
 had a small impact on the overall
 chord estimation accuracies.

\begin{figure*}[t]
\centering
 \includegraphics[width=2.0\columnwidth]{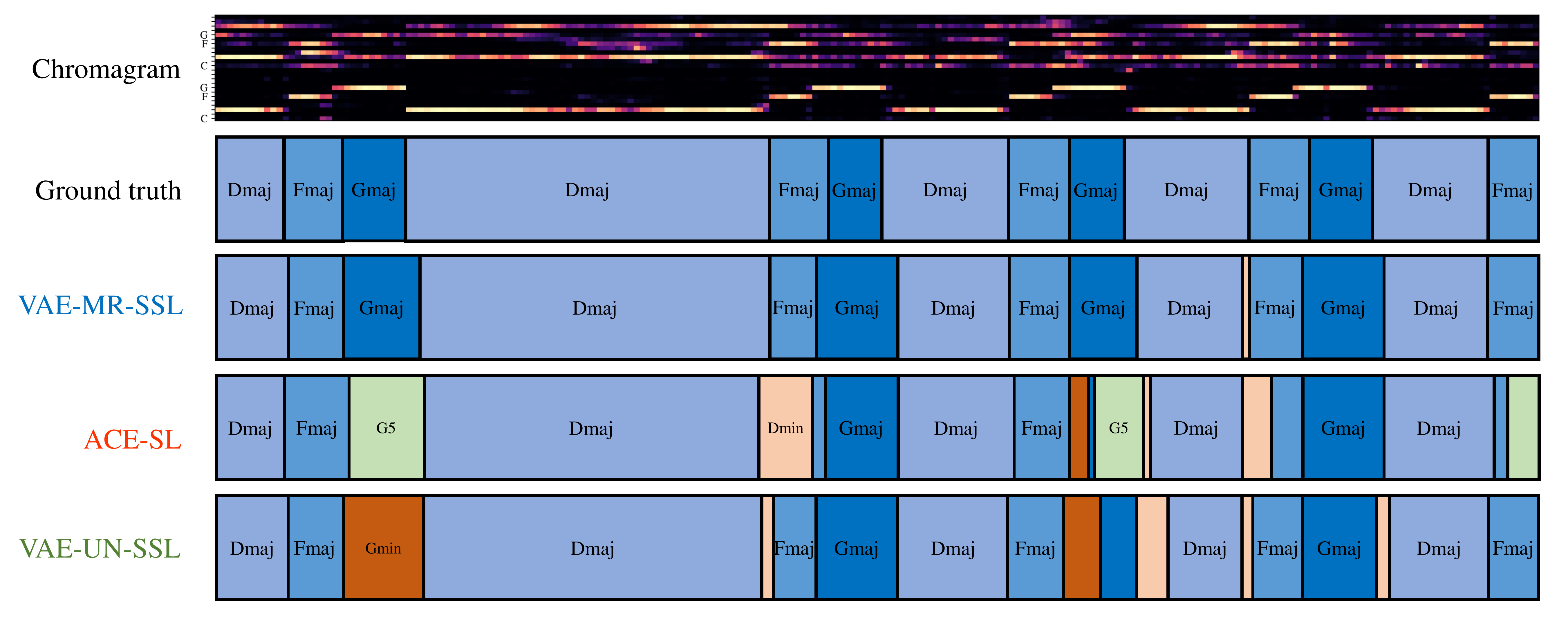}
  \vspace{-2mm}
 \caption{
 An example of chord label sequences
 estimated by the supervised and semi-supervised methods
 without the Viterbi post-filtering. 
 For readability, only the first 24 dimensions (bass and middle channels) of the chroma vectors are displayed.}
 \label{fig:estimations}
\end{figure*} 

In Fig.~\ref{fig:estimations},
 some errors made by \textbf{ACE-SL} (light-green segments)
 were corrected by the VAE-based methods.
The chord label sequence obtained by \textbf{VAE-UN-SSL}, however,
 included a number of incorrect short fragments
 caused by the local fluctuations of the chroma vectors.
As discussed above,
 $q_\alpha(\mS|\mX)$ was encouraged to frequently vary over time
 such that the fine structure of chroma vectors 
 could be reconstructed precisely in the VAE framework.
In contrast, the chord label sequence obtained by \textbf{VAE-MR-SSL} 
 included fewer transitions
 and was much closer to the ground-truth sequence.

\begin{figure}[t]
 \centering
 \centerline{\includegraphics[width=.80\columnwidth]{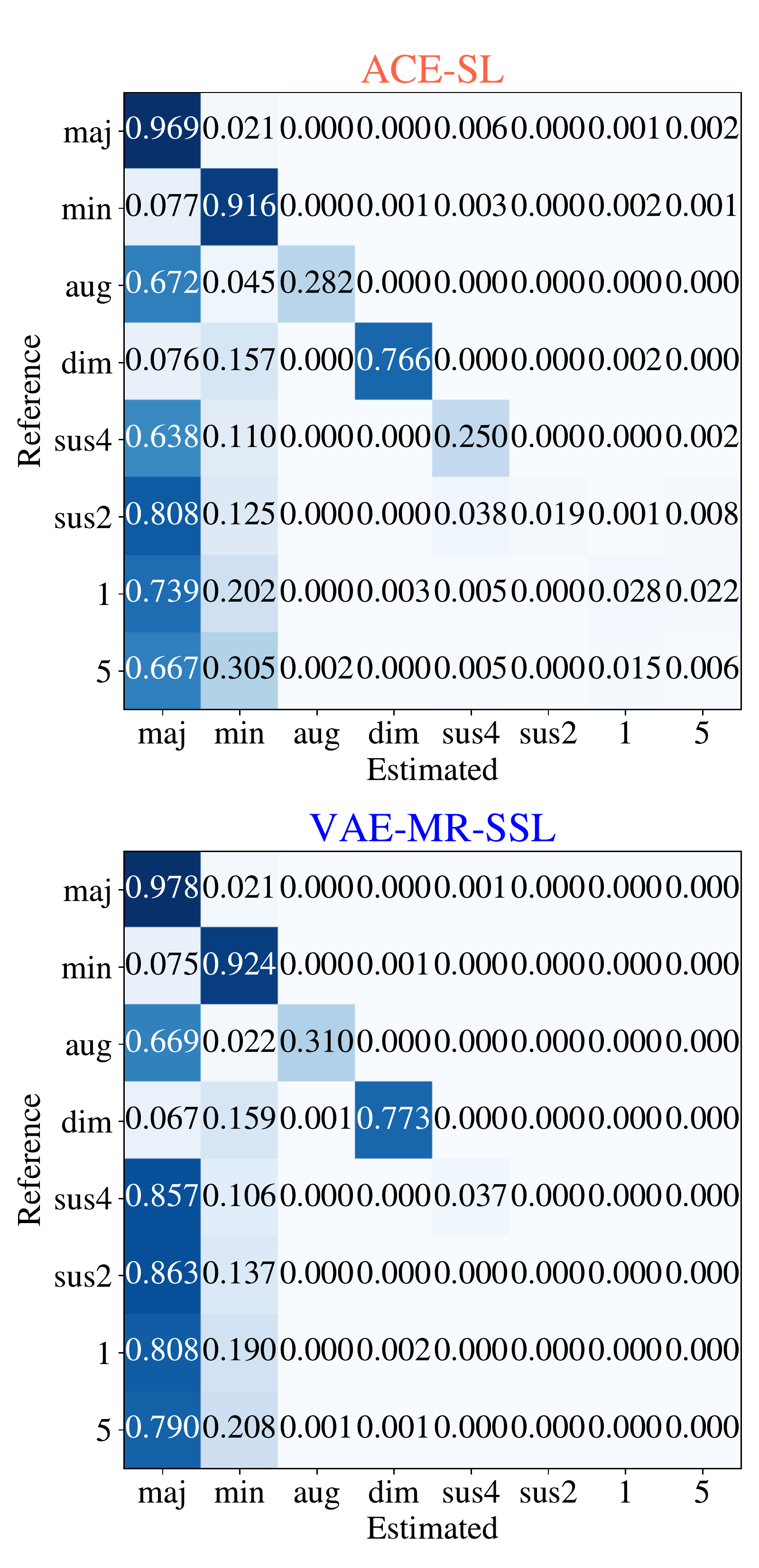}}
 \vspace{-2mm}
 \caption{Confusion matrices with respect to chord types. 
 Estimated chords with wrong root notes are not counted.}
 \label{fig:confmatrix}
\end{figure} 

\begin{figure}[t]
 \centering
 \centerline{\includegraphics[width=0.75\columnwidth]{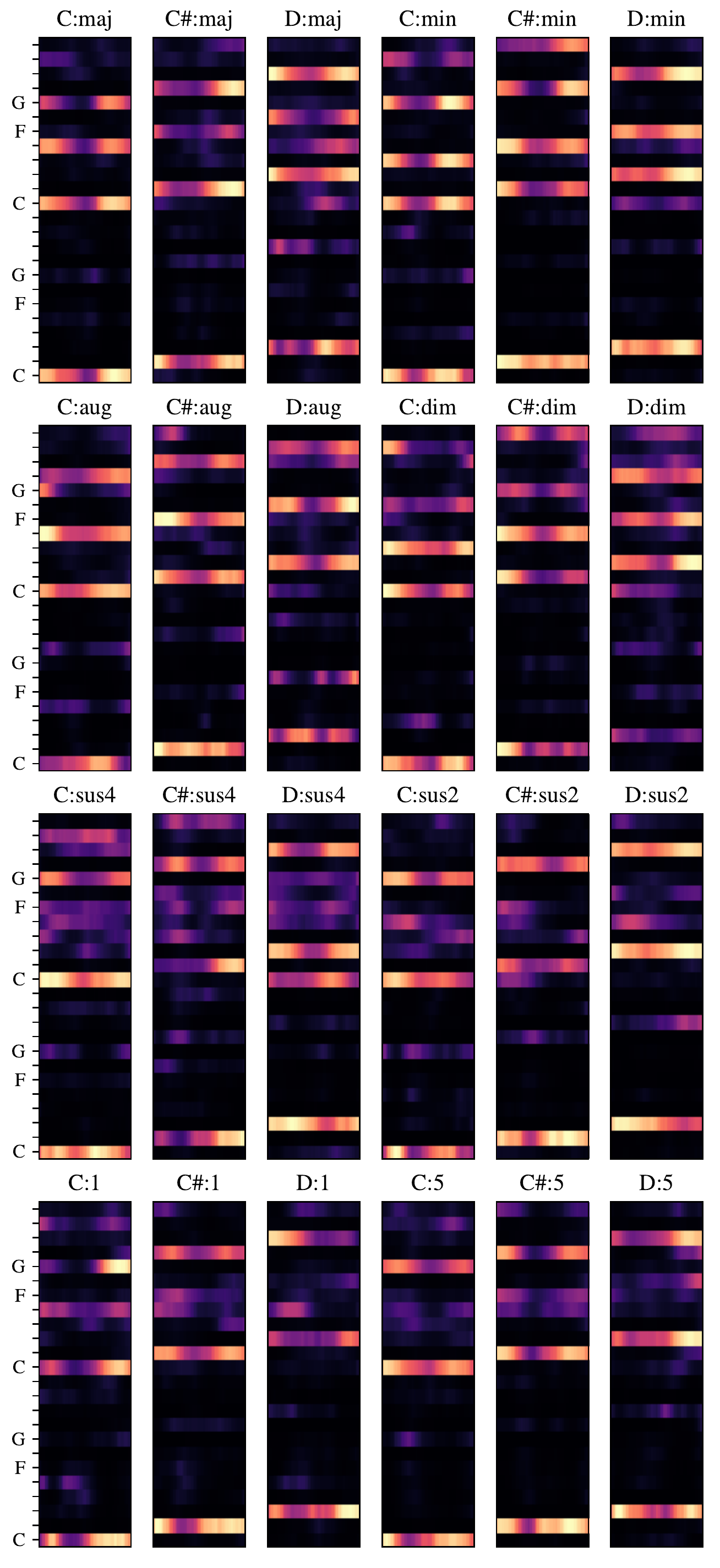}}
 \caption{The probability distributions obtained by $p_\theta(\mX|\mS,\mZ)$ conditioned 
  by different chord labels $\mS$. 
  Only the first 24 dimensions (bass and middle channels) of the chroma vectors are displayed.}
 \label{fig:analogies}
\end{figure} 

\subsection{Further Observations}

Fig.~\ref{fig:confmatrix} shows
 the confusion matrices obtained by \textbf{ACE-SL} and \textbf{VAE-MR-SSL}.
While the accuracies on the \textit{maj}, \textit{min}, \textit{aug} and \textit{dim} types 
 were improved by \textbf{VAE-MR-SSL}, 
 the accuracies on the other uncommon chord types were significantly degraded.
Rare chords tended to be wrongly classified to
 the \textit{maj} and \textit{min} triads.
Interestingly, the accuracies on the eight chord types 
 were not necessarily correlated to their ratios
 in the training data (Table \ref{tab:shorthands}),
 where the \textit{sus4} triads
 were much more frequently used than the \textit{aug} and \textit{dim} triads.
The same problem occurred in \textbf{VAE-UN-SSL}
 because the unsupervised learning objective (\ref{eq:objective_unsupervised})
 has no mechanism to prevent $q_\alpha(\mS|\mX)$ from excessively yielding popular chord types.

As shown in Fig.~\ref{fig:analogies}, 
 the generative model $p_\theta(\mX|\mS,\mZ)$
 successfully reconstructed chroma vectors,
 conditioned by the \textit{maj}, \textit{min}, \textit{aug}, and \textit{dim} triads,
 \ie, the reconstructed chroma vectors
 had high probabilities on the pitch classes of the chord notes.
In contrast, $p_\theta(\mX|\mS,\mZ)$
 failed to reconstruct chroma vectors,
 when conditioned by
 the \textit{sus2} and \textit{sus4} triads and the power chords.
Comparing Fig. \ref{fig:confmatrix} with Fig.~\ref{fig:analogies},
 we investigate the relationships between
 the estimation accuracy of chord labels
 and the reconstruction quality of chroma vectors.
The generator $p_\theta(\mX|\mS,\mZ)$ failed to
 learn the pitch-class distributions 
 of several chord classes (\eg, \textit{sus2} and \textit{sus4})
 whose chroma vectors often had no clear peaks on the chord notes.
For such chroma vectors, 
 $p_\theta(\mX|\mS,\mZ)$ always gave a lower probability
 even if $\mS$ was the ground-truth chord classes corresponding to $\mX$.
Note that $p_\theta(\mX|\mS,\mZ)$ was used 
 for regularizing the classifier $q_\alpha(\mS|\mX)$
 in the unsupervised learning objective (\ref{eq:objective_unsupervised}),
 where $q_\alpha(\mS|\mX)$ was trained to
 avoid \textit{sus2} and \textit{sus4} triads 
 and favor \textit{maj} triads to maximize $p_\theta(\mX|\mS,\mZ)$.

\section{Conclusion}
\label{sec:conclusion}

This paper described a statistical method
 that trains a neural chord estimator in a semi-supervised manner
 by constructing a VAE with latent chord labels and features.
This is a new approach to ACE
 that unifies the generative and discriminative methods.
Our method incorporates 
 a Markov prior on chord labels to encourage the temporal continuity of chord labels
 estimated with the chord estimator.
The comparative experiment clearly showed 
 the effectiveness of using the generative model of chroma vectors
 and the Markov prior on chord labels
 in regularizing the chord estimator for performance improvement.
The limitation of the proposed semi-supervised learning
 is that the trained chord classifier tends to 
 mistakenly classify some specific chord types into popular types.
This may become more problematic 
 when a larger chord vocabulary including
 seventh chords and chord inversions is used.

The success of the semi-supervised VAE for ACE
 indicates the effectiveness of unifying 
 the deep generative and discriminative methods 
 in automatic music transcription (AMT).
Using the AVI framework,
 we can explicitly
 introduce prior knowledge on musical symbol sequences
 as a regularization term.
Such knowledge is hard to automatically extract from training data
 in supervised discriminative methods.
One way to improve the performance of ACE
 is to replace the frame-level Markov prior of chord labels
 with a beat- or symbol-level language model,
 which has been considered to be effective for
 solving the ambiguity in acoustic features \cite{korzeniowski2017on,Korzeniowski2018}.
We also plan to develop a comprehensive AMT system
 based on a unified VAE that can treat mutually-dependent musical elements
 such as keys, beats, and notes as latent variables.

\bibliographystyle{IEEEtran}
\bibliography{refs}

\end{document}